\documentclass[a4paper,11pt]{article}
\usepackage{amssymb,graphicx,amsmath}

\makeatletter 
\renewcommand\section{\@startsection{section}{1}{\z@}              {-3.25ex\@plus -1ex \@minus -.2ex}                                    {1.5ex \@plus .2ex}                                    {\reset@font\Large\bfseries}}
\renewcommand\subsection{\@startsection{subsection}{2}{\z@}                                    {3.25ex \@plus 1ex \@minus.2ex}                                    {-1em}                                    {\reset@font\large\bfseries}}
\@addtoreset{equation}{section} \makeatother
\renewcommand{\theequation}{\thesection.\arabic{equation}}
\input{tcilatex}

\newcommand{\bc}{\begin{center}}
\newcommand{\ec}{\end{center}}
\def\ba#1{\begin{array}{#1}\displaystyle}
\newcommand{\ea}{\end{array}}
\newcommand{\z}{\\[2mm] \displaystyle}

\newcommand{\beq}{\begin{equation}}
\newcommand{\eeq}{\end{equation}}
\newcommand{\beqa}{\begin{eqnarray}}
\newcommand{\eeqa}{\end{eqnarray}}

\newcommand{\n}{\nonumber\\}
\newcommand{\bi}{\begin{itemize}}
\newcommand{\ei}{\end{itemize}}

\def\lt#1{\left#1}
\def\rt#1{\right#1}
\def\t#1{\tilde{#1}}
\def\h#1{\hat{#1}}
\def\b#1{\bar{#1}}
\def\frc#1#2{\frac{#1}{#2}}

\newcommand{\p}{\partial}

\newcommand{\vac}{{\rm vac}}
\newcommand{\bra}{\langle}
\newcommand{\ket}{\rangle}
\newcommand{\Z}{{\mathbb{Z}}}
\newcommand{\N}{{\mathbb{N}}}

\newcommand{\Or}{{\cal O}}

\newcommand{\ep}{\epsilon}
\newcommand{\varep}{\varepsilon}

\newcommand{\Tr}{{\rm Tr}}

\begin{document}

\setcounter{page}{0} \topmargin0pt \oddsidemargin0mm \renewcommand{%
\thefootnote}{\fnsymbol{footnote}} \newpage \setcounter{page}{0}
\begin{titlepage}

\vspace{0.2cm}
\begin{center}
{\Large {\bf Bi-partite entanglement entropy in integrable models with
backscattering}}

\vspace{0.8cm} {\large  \text{ O.A.~Castro-Alvaredo$^{\bullet}$
and B.~Doyon$^{\circ}$}}

\vspace{0.2cm}{$^{\bullet}$  Centre for Mathematical Science, City University London, \\
Northampton Square, London EC1V 0HB, UK} \\
\vspace{0.2cm}
{$^{\circ}$  Department of Mathematical Sciences, Durham University \\ South Road, Durham DH1 3LE, UK }
\end{center}
\vspace{1cm}

In this paper we generalise the main result of a recent work by
J.~L.~Cardy and the present authors concerning the bi-partite entanglement
entropy between a connected region and its complement. There the expression of the
leading order correction to saturation in
the large distance regime was obtained for integrable quantum
field theories possessing diagonal scattering matrices. It was
observed to depend only on the mass spectrum of the model and not
on the specific structure of the diagonal scattering matrix. Here
we extend that result to integrable models with backscattering
(i.e. with non-diagonal scattering matrices). We use again the replica method,
which connects the entanglement entropy to partition functions on Riemann surfaces with
two branch points. Our main conclusion
is that the mentioned infrared correction takes exactly the same
form for theories with and without backscattering. In order to
give further support to this result, we provide a detailed
analysis in the sine-Gordon model in the coupling regime in which
no bound states (breathers) occur. As a consequence, we obtain the leading correction
to the sine-Gordon partition function on a Riemann surface in the large distance regime.
Observations are made concerning the limit of large number of sheets.

\vfill{ \hspace*{-9mm}
\begin{tabular}{l}
\rule{6 cm}{0.05 mm}\\
$^\bullet \text{o.castro-alvaredo@city.ac.uk}$\\
$^\circ \text{benjamin.doyon@durham.ac.uk}$\\
\end{tabular}}

\renewcommand{\thefootnote}{\arabic{footnote}}
\setcounter{footnote}{0}

\end{titlepage}
\newpage

\section{Introduction}

A quantity of current interest in quantum models with many local
degrees of freedom is the bi-partite entanglement entropy
\cite{bennet}. It is a measure of quantum entanglement between the
degrees of freedom of two regions, $A$ and its complement, in the
ground state $|{\rm gs}\ket$ of the model. Other measures of
entanglement also exist, see e.g. \cite{bennet}-\cite{Verstraete}.
Such measures occur in the context of quantum computing, for
instance. Since entanglement is a fundamental property of quantum
systems, a measure of entanglement gives a good description of the
quantum nature of a ground state, perhaps more so than correlation
functions. For the formal definition of the entanglement entropy,
consider the Hilbert space as a tensor product of local spaces
associated to the sites of a quantum system. This can be written
as a tensor product of the two quantum spaces associated to the
regions $A$ and its complement: ${\cal H} = {\cal A} \otimes
\b{{\cal A}}$. Then the entanglement entropy is the von Neumann
entropy of the reduced density matrix $\rho_A$ associated to $A$:
\beq
    S_A = \Tr_{{\cal A}} \rho_A\log \rho_A ~,\quad \rho_A = \Tr_{\b{{\cal A}}} |{\rm gs}\ket \bra {\rm gs}|~.
\eeq

In this work we will be interested in 1-dimensional quantum
systems. The entanglement of quantum spin chains has been
extensively studied in the literature \cite{Eisert}-\cite{Weston}.
The scaling limit, describing the universal part of the quantum
chain behaviour near a quantum critical point, is a quantum field
theory (QFT) model (which we will assume throughout to possess
Poincar\'e invariance). The scaling limit is obtained by
approaching the critical point while letting the length of the
region $A$ go to infinity in a fixed proportion with the
correlation length. It is known since some time
\cite{CallanW94}-\cite{Calabrese:2005in} that the bi-partite
entanglement entropy can be re-written in terms of more geometric
quantities in this limit, using a ``replica trick''. It is related
to the partition function $Z_n(x_1,x_2)$ of the (euclidean) QFT
model on a Riemann surface ${\cal M}_{n,x_1,x_2}$ with two branch
points, at the points $x_1$ and $x_2$, and $n$ sheets cyclicly
connected. The position of the branch points correspond to the
end-points of the region $A$ in the scaling limit. The relation is
based on the simple re-writing $S_A = -\lim_{n\to 1}\frc{d}{dn}
\Tr_{{\cal A}} \rho_A^n$, which gives: \beq
    S_A = -\lim_{n\to 1}\frc{d}{dn} \frc{Z_n(x_1,x_2)}{Z_1^n}~.
\eeq Naturally, this expression implies that we must  analytically
continue the quantity $Z_n(x_1,x_2)$ from $n\in \N$, where it is naturally
associated to Riemann surfaces, to
$n\in[1,\infty)$. The object $\Tr_{{\cal A}} \rho_A^n$ certainly has a well-defined meaning
for any $n$ such that ${\rm Re}(n)>0$. Indeed, $\rho_A$ is hermitian (and has non-negative eigenvalues summing to 1), so that
$\Tr_{{\cal A}} \rho_A^n$ is the sum of the $n^{\rm th}$ powers of its eigenvalues (with multiplicities). Note that this is an
analytic continuation from positive integers $n$ to complex $n$ that satisfies the requirements of Carlson's theorem \cite{Rubel55}, hence the unique one that does.
The scaling limit of this object is what defines the proper analytic continuation of $Z_n(x_1,x_2)$.
It is natural to assume, as it has been done before \cite{CallanW94}, that the two branch points
just become conical singularities with angle $2\pi n$, the rest of the space being flat. This is the point of view that we will take.

In \cite{entropy}, the ratio of partition  functions
$Z_n(x_1,x_2)/Z_1^n$ was studied at large distances $|x_1-x_2|=r$
for certain 1+1-dimensional integrable QFTs. The main feature of
these models is that there is no particle production in any
scattering process and the scattering ($S$) matrix factorizes into
products of 2-particle $S$-matrices which can be calculated
exactly (for reviews see e.g.
\cite{Karowski:1978eg}-\cite{Dorey:1996gd}). Taking the $S$-matrix
as input it is then possible to compute the matrix elements of
local operators (also called form factors). This is done by
solving a set of consistency equations \cite{KW,Smirnovbook}. This
is known as the form factor bootstrap program for integrable QFTs.
In \cite{entropy}, this program was used and generalised in order
to compute $Z_n(x_1,x_2)/Z_1^n$ in the case of integrable models
with diagonal scattering matrix (that is, without backscattering).
It was deduced that for this class of models, the entanglement
entropy behaves at large length $r$ of the region $A$ as
\beq\label{main}
    S_A =-\frac{c}{3}\log(\varepsilon m_1) -  \frc18 \sum_{\alpha=1}^\ell K_0(2rm_\alpha) + O\lt(e^{-3rm_1}\rt).
\eeq Here, $m_\alpha$ are the masses of the $\ell$ particles in
the QFT model,  with $m_1\leq m_\alpha$. The first term is the
expected saturation of the entanglement entropy (with $c$ the
ultraviolet central charge), the precise value of which depends on
the details of the microscopic theory included into the
non-universal small distance $\varep$. The interesting feature is
the universal second term, where we see that the leading
exponential corrections are independent of the scattering matrix,
and only depend on the particle spectrum of the model. This is
quite striking: for instance, a model of $\ell$ free particles of
masses $m_\alpha$ will give the same leading exponential
corrections as one with diagonally interacting particles with the
same masses.

The purpose of this paper is to point out that this result still holds
for {\em non-diagonal} integrable models, and to analyse the particular
case of the sine-Gordon model. We obtain the first large-distance correction
to $Z_n(x_1,x_2)/Z_1^n$ for that model and provide a
detailed analysis of its large-$n$ behaviour. This analysis gives
a numerical confirmation of the main result (\ref{main}).
The large-$n$ behaviour also shows interesting features that are in
relation with the properties of the perturbing field: it is linear $\propto n$
in the super-renormalisable case, and has an extra logarithmic factor
$\propto n\log n$ in the marginally renormalisable case (where the sine-Gordon
model represents a sector of the $SU(2)$-Thirring model).

The paper is organised as follows: In section 2 we review the
relationship, discussed in \cite{entropy}, between the partition function (hence the
entanglement entropy) and the two-point function of branch-point twist
fields in a theory consisting of $n$ copies of a given integrable
model. We generalise the procedure introduced in \cite{entropy} for
computing the entanglement entropy from analytic properties of form
factors of branch-point twist fields to integrable models including backscattering.
This is based on expressing the two-point function in terms of a
form factor expansion. For all such models we conclude that, under certain
assumptions, the leading order correction to the entropy in the
infrared limit is of the same form as that obtained in
\cite{entropy} for theories with diagonal scattering. In section 3
we introduce the sine-Gordon model and obtain the two-particle
form factors of the branch-point twist fields. In
section 4 we check these form factors for consistency by
extracting the underlying conformal dimension of the twist fields.
In section 5 we check our general result from
section 2 for the sine-Gordon model, and provide a large-$n$ analysis
of the two-particle contribution to the two-point function. We find good analytical and
numerical agreement with the results of section 2. In section 6 we
present our conclusions and outlook.

\section{Form factor equations and the entanglement entropy}

\subsection{Form factor equations}\indent \\
\indent This section follows closely \cite{entropy}. Let us
consider some model of 1+1-dimensional QFT. For clarity, we will
characterise it by its lagrangian density ${\cal L}(\phi)$
depending on some ``fundamental'' field $\phi$, although the
results hold also when no lagrangian formulation is available. In
\cite{entropy}, it was shown that the partition function
$Z_n(x_1,x_2)$ on ${\cal M}_{n,x_1,x_2}$ is proportional to a
two-point correlation function of certain twist fields in an
extended model, composed of $n$ independent copies of the initial
theory. The lagrangian density of this extended model, for
instance, is ${\cal L}^{(n)}(\phi_1,\ldots,\phi_n) = \sum_{j=1}^n
{\cal L}(\phi_j)$. This model possesses a natural $\Z_n$ internal
symmetry under cyclic exchange of the copies, $\sigma:
\phi_j\mapsto \phi_{j+1}~ (j=1,\ldots,n-1),~\phi_{n}\mapsto
\phi_1$. By a standard procedure (see, for instance, the
explanations in \cite{entropy}), to this symmetry and to that of
the opposite cyclic exchange $\sigma^{-1}$, we can associate,
respectively, the twist fields ${\cal T}$ and $\t{{\cal T}}$,
called {\em branch-point twist fields}. Their two-point function
is simply related to the partition function of the original model
on the Riemann surface: \beq
    \frc{Z_n(x_1,x_2)}{Z_1^n} = {\cal Z}_n \varep^{2d_n}\bra {\cal T}(x_1) \t{\cal T}(x_2) \ket
\eeq
where here an below we use $\bra\cdots\ket$ for denoting correlation functions in the extended
 model ${\cal L}^{(n)}$. Here $\varep$ is some short-distance scale, ${\cal Z}_n$ is a non-universal
  normalisation (with ${\cal Z}_1=1$), and $d_n$ is the scaling dimension of the twist fields \cite{Calabrese:2004eu,entropy}
\beq\label{dim}
    d_n = \frc{c}{12} \lt(n-\frc1n\rt),
\eeq where $c$ is the central charge of the ultra-violet conformal
field theory associated to ${\cal L}$ (that describes the
short-distance behaviours of correlation functions in the model
${\cal L}$). The fundamental property of the twist fields ${\cal
T}$ and $\t{\cal T}$, as operators on the Hilbert space of ${\cal
L}^{(n)}$, is the ``semi-locality'' property with respect to any
local field $\Or$: \beqa
    \Or(y) {\cal T} (x) &=& {\cal T}(x) (\sigma\Or)(y) \quad x^1<y^1 \n
    \Or(y) {\cal T} (x) &=& {\cal T}(x) \Or(y) \quad x^1>y^1 \n
    \Or(y) \t{\cal T} (x) &=& \t{\cal T}(x) (\sigma^{-1}\Or)(y) \quad x^1<y^1 \n
    \Or(y) \t{\cal T} (x) &=& \t{\cal T}(x) \Or(y) \quad x^1>y^1~.
\label{semiloc} \eeqa This implies that $\Or$ and ${\cal T}$ (or
$\t{\cal T}$) are mutually  local\footnote{That is, they commute,
hence are quantum mechanically independent, at space-like
distances.} only when $\Or$ is invariant under $\sigma$. Along
with the fact that they have the lowest dimension, given by
(\ref{dim}), and that they are invariant under all other symmetry
transformations of the model ${\cal L}^{(n)}$ that are in
agreement with $\sigma$, this fixes ${\cal T}$ and $\t{\cal T}$
uniquely up to normalisation (for definiteness, we will assume the
CFT normalisation, $\bra{\cal T}(x) \t{\cal T}(0) \ket \sim
|x|^{-2d_n}$).

An important remark is that these twist fields are local fields:
they commute  with the energy density at space-like
distances\footnote{The set of all local fields that constitute a
QFT is the set of those that are mutually local with respect to the energy
density.}. This is a consequence of the fact that they are
associated to a symmetry. In particular, the resulting two-point
function, which is proportional to the partition function of
${\cal L}^{(n)}$ with a defect line extending from $x_1$ to $x_2$
through which the fields are affected by the symmetry
transformation, is independent of the shape of this defect line
(this is sometimes called a topological defect). This is simply
related to the fact that the partition function of ${\cal L}$ on
the Riemann surface ${\cal M}_{n,x_1,x_2}$ is independent from the
shape of the branch connections.

Let us now consider ${\cal L}$ to be a massive integrable QFT
model. The Hilbert  space of massive QFT is described by physical
particles, and there are two bases, one corresponding to
well-defined separated particles coming from far in the past, the
other corresponding to those leaving far in the future. The
overlap between these bases is the scattering matrix, and the main
characteristic of integrable models is that their scattering
preserves the number of particles involved and the set of momenta,
and that it factorises into two-particle processes, as if
particles were scattering by pairs at very distant space-time
points from each other. Hence, only the two-particle to
two-particle scattering matrix is relevant, and consistency under
all possible pair-wise separations gives the Yang-Baxter equation.
Consider an integrable model with mass spectrum
$\{m_\alpha,\;\alpha=1,\ldots,\ell\}$ (some masses may be equal).
Let us denote by
$|\theta_1,\ldots,\theta_k\ket_{\alpha_1\ldots\alpha_k}$ with
$\theta_1>\cdots>\theta_k$ the in-states with $k$ particles of
rapidities $\theta_1,\ldots,\theta_k$ and, respectively, of
particle types (quantum numbers) $\alpha_1,\ldots,\alpha_k$. For
$\theta_1<\ldots<\theta_k$, this will represent the out-states.
Then, the two-particle scattering matrix is defined by
\[
    |\theta_1,\theta_2\ket_{\alpha_1\alpha_2} =
    \sum_{\beta_1,\beta_2} S_{\alpha_1\alpha_2}^{\beta_1\beta_2}(\theta_1-\theta_2) |\theta_2,\theta_1\ket_{\beta_2\beta_1}
\]
where we used relativistic invariance in order to write the
scattering matrix  as function of the rapidity differences. This
equation holds for $\theta_1>\theta_2$ as well as
$\theta_1<\theta_2$, thanks to unitary. One of the main
achievements of massive integrable QFT is the exact calculation of
scattering matrices in many models, from its expected physical
properties along with the Yang-Baxter equation (or, in some cases,
from Bethe ansatz solution of an underlying integrable microscopic
model).

Another important result of massive integrable QFT is the exact
calculation  of some form factors in many models
\cite{KW,Smirnovbook} (see \cite{Essler:2004ht} for a recent
review). Form factors are matrix elements of local fields between
the vacuum and a many-particle state (say an in-state): \beq
    F_k^{\Or|\alpha_1\ldots\alpha_k}(\theta_1,\ldots,\theta_k) = \bra \vac|\Or(0) |\theta_1,\ldots,\theta_k\ket_{\alpha_1\ldots\alpha_k}~.
\eeq They are, more precisely, the analytic continuation of these
matrix elements  in the rapidity variables. The exact evaluation
of these objects follows from solving a set of expected properties
that involve the exact two-particle scattering matrix and form a
Riemann-Hilbert problem, along with certain minimality assumptions
on the analytic structure. Besides integrability, an important
requirement for this Riemann-Hilbert problem to hold is locality
of the field $\Or$. Indeed, a strong indication of its validity is
that it is possible to show that two fields whose form factors
solve it commute at space-like distances. For the simplest local
fields, this Riemann-Hilbert problem is well known and can be
solved in many cases.

If ${\cal L}$ is integrable, then certainly ${\cal L}^{(n)}$ also
is, with $n$ times as  many particles, which we will denote by the
double index $(\alpha,i)$ for $\alpha=1,\ldots,\ell$ and
$i=1,\ldots,n$. Its scattering matrix is simply given by
\beq\label{scat}
    S_{(\alpha_1,i_1)\,(\alpha_2,i_2)}^{(\beta_1,j_1)\,(\beta_2,j_2)}(\theta) = \delta_{i_1,j_1} \delta_{i_2,j_2}\times \lt\{
        \ba{ll} \delta_{\alpha_1,\beta_1} \delta_{\alpha_2,\beta_2} & i_1\neq i_2 \z
        S_{\alpha_1\alpha_2}^{\beta_1\beta_2}(\theta) & i_1=i_2 \ea \rt. ~.
\eeq Locality of the twist fields ${\cal T}$ and $\t{\cal T}$
along with the exchange relations (\ref{semiloc})  were used in
\cite{entropy} in order to justify a Riemann-Hilbert problem for
their form factors in integrable models with diagonal scattering
(that is, with $S_{\alpha_1\alpha_2}^{\beta_1\beta_2}(\theta) =
S_{\alpha_1\alpha_2}(\theta) \delta_{\alpha_1,\beta_1}
\delta_{\alpha_2,\beta_2}$). The generalisation to non-diagonal
scattering is straightforward. We will consider only ${\cal T}$,
since $\t{\cal T} = {\cal T}^\dag$ on the Hilbert space. Let us
denote by $\mu$, $\nu$ and $\omega$ double indices of the form
$(\alpha,j)$, and by $F_{k}^{\mathcal{T}|\mu_1\ldots\mu_k
}(\theta_1,\ldots, \theta_{k},n )$ the form factors of ${\cal T}$
in the $n$-copy model ${\cal L}^{(n)}$. They are analytic in the
rapidity differences except for poles (that may accumulate at
infinity), and satisfy the relations
\begin{eqnarray}
  F_{k}^{\mathcal{T}|\ldots \mu_i \mu_{i+1} \ldots }(\ldots,\theta_i, \theta_{i+1}, \ldots,n ) &=&
  \sum_{\nu_i,\nu_{i+1}} S_{\mu_i \mu_{i+1}}^{\nu_i \nu_{i+1}} (\theta_{i\,i+1})
  F_{k}^{\mathcal{T}|\ldots \nu_{i+1}  \nu_{i} \ldots}(\ldots,\theta_{i+1}, \theta_i,  \ldots,n ), \n
 F_{k}^{\mathcal{T}|\mu_1 \mu_2 \ldots \mu_k}(\theta_1+2 \pi i, \ldots,
\theta_k,n) &=&
  F_{k}^{\mathcal{T}| \mu_2 \ldots \mu_k \hat{\mu}_1}(\theta_2, \ldots, \theta_{k},
  \theta_1,n), \label{qpt}
  \end{eqnarray}
  and
\begin{eqnarray}
-i \text{Res}_{\substack{\bar{\theta}_{0}={\theta}_{0}}}
 F_{k+2}^{\mathcal{T}|\b{\mu} \mu  \mu_1 \ldots \mu_k}(\bar{\theta}_0+i\pi,{\theta}_{0}, \theta_1 \ldots, \theta_k,n)
  &=&
  F_{k}^{\mathcal{T}| \mu_1 \ldots \mu_k}(\theta_1, \ldots,\theta_k,n), \n
 -i \text{Res}_{\substack{\bar{\theta}_{0}={\theta}_{0}}}
 F_{k+2}^{\mathcal{T}|\b\mu \hat{\mu } \mu_1 \ldots \mu_k}(\bar{\theta}_0+i\pi,{\theta}_{0}, \theta_1 \ldots, \theta_k,n)
  &=&\sum_{\nu_1,\ldots,\nu_k} \mathcal{P}_{\hat{\mu} \mu_1 \ldots \mu_k}^{\nu_1,\ldots,\nu_k}(\theta_0, \theta_1, \ldots, \theta_k,n)
  \n &&\times
  F_{k}^{\mathcal{T}| \nu_1 \ldots \nu_k}(\theta_1, \ldots,\theta_k,n)\label{kre}
\end{eqnarray}
with
\begin{equation}
    \mathcal{P}_{\hat{\mu} \mu_1 \ldots \mu_k}^{\nu_1,\ldots,\nu_k}(\theta_0, \theta_1, \ldots,
    \theta_k,n)=
  \sum_{\omega_1,\ldots,\omega_{k-1}}
    S_{\hat{\mu}\mu_1}^{\omega_1 \nu_1}(\theta_{01})
    S_{\omega_1 \mu_2}^{\omega_2 \nu_2}(\theta_{02})\cdots S_{\omega_{k-1} \mu_k}^{\h\mu
    \nu_k}(\theta_{0k}).
\end{equation}
Here $\theta_{ij}=\theta_i-\theta_j$. As function of $\theta_1$
for real $\theta_2,\ldots,\theta_k$,  there are no poles of
$F_{k}^{\mathcal{T}|\mu_1\ldots\mu_k }(\theta_1,\ldots,
\theta_{k},n )$
in the strip ${\rm Im}(\theta_1) \in [0,\pi]$, except for those
given by the last two equations, and for poles at
purely imaginary values (and with purely imaginary residues) corresponding to bound states between the
associated particles (note that we will not consider any bound states in the
sine-Gordon example studied below).
In the second equation, the crossing or locality
relation, we introduced the symbol $\hat{\mu}=(\alpha,j+1)$. As
compared to the usual form factor equations, it is altered by the
nature of the exchange relation and it now relates form factors
associated to different particle sets (belonging to different
copies). Finally, the last two equations generalise the standard
kinematic residue equation to branch-point twist fields, where we
introduced the symbol $\b\mu=(\b\alpha,j)$ with $\b\alpha$
denoting the quantum number of the anti-particle of $\alpha$ in
the theory ${\cal L}$. Once more, the exchange relations
(\ref{semiloc}) are responsible for the splitting into two
equations, and the shift in $\hat{\mu}$.

It is instructive to specialise to two particles;
this is what gives the main result (\ref{main}) for the entropy
(one-particle form factors of spinless fields are $\theta$-independent). In this case, the first two form factor equations specialise to
\begin{eqnarray}
F_{2}^{\mathcal{T}|(\alpha,j)\,(\beta,k)}(\theta,
n)=\sum_{\delta,\gamma} F_{2}^{\mathcal{T}|(\gamma,k)\, (\delta,j) }(-\theta,
n)S_{(\alpha,j)\, (\beta,k) }^{(\delta,j) \,(\gamma,k) }(\theta)=
F_{2}^{\mathcal{T}|( \beta,k)\, ( \alpha,j+1)}(2\pi i-\theta, n),
\label{minix}
\end{eqnarray}
for all values of $j$, $k$, $\alpha$ and $\beta$, where $\theta$ is now the rapidity difference. From the equations
above, from (\ref{scat}) and from application of the $\Z_n$ symmetry, it follows:
\begin{eqnarray}
  F_{2}^{\mathcal{T}|(\alpha,i) \,(\beta,i+k)}(\theta,n) &=&
  F_{2}^{\mathcal{T}|(\alpha,j)\,  (\beta, j+k) \,}(\theta,n) \qquad
\forall \quad i,j,k,\alpha,\beta\label{usex}\\
F_{2}^{\mathcal{T}|(\alpha,1)\,(\beta,j)}(\theta,
n)&=&F_{2}^{\mathcal{T}|(\beta,1)\,(\alpha,1)}(2\pi(j-1)i-\theta, n
)\quad \forall \quad \alpha,\beta,j\in\{2,\ldots,n+1\}.\label{use2x}
\end{eqnarray}
The last equation at $j=n+1$, and the first equation of (\ref{qpt}), give
\begin{equation}\label{maineq}
F_{2}^{\mathcal{T}|(\alpha,1)\,(\beta,1)}(\theta,
n)=\sum_{\delta,\gamma} F_{2}^{\mathcal{T}|(\gamma,1)\, (\delta,1) }(-\theta,
n)S_{\alpha\beta }^{\delta\gamma }(\theta)=
F_{2}^{\mathcal{T}|( \beta,1)\, ( \alpha,1)}(2\pi n i-\theta, n),
\end{equation}
so, as in \cite{entropy}, we can solve for the form factors associated to
particles in the first copy and then use (\ref{use2x}) to
obtain all other solutions. From now on we will abbreviate
\begin{equation}
  F_{2}^{\mathcal{T}|(\alpha,1)(\beta,1)}(\theta, n) := F_{2}^{\mathcal{T}|\alpha \beta}(\theta, n).
\end{equation}

The simple form of the scattering matrix for particles living in
different sheets, and the fact that no bound state can occur
between particles in different copies, simplifies drastically the
pole structure of the form factors. A combination of the equations
above along with the two-particle case of the last two equations
of (\ref{kre}) gives that $F_2^{{\cal T}|\alpha\beta}(\theta,n)$
is analytic for ${\rm Im}(\theta)\in[0,2\pi n]$ except if
$\b\alpha = \beta$ for poles at $\theta=i\pi$, with residue
$i\bra{\cal T}\ket$, and at $\theta=2i\pi n-i\pi$, with residue
$-i\bra{\cal T}\ket$, and for possible poles in ${\rm
Im}(\theta)\in(0,\pi)$ and in ${\rm Im}(\theta)\in(2\pi n-\pi,2\pi
n)$ corresponding to bound states.

\subsection{Conical singularities and the entanglement
entropy}\indent\\
\indent The analytic structure described above is sufficient to
establish the result (\ref{main}), following arguments of
\cite{entropy}. The most delicate point of these arguments is the
analytic continuation in $n$, which still needs further
justification. Below we attempt to support the arguments from the
geometrical picture of conical singularities. Another point is
that we must assume that form factors at real rapidities vanish
faster than $(n-1)^{\frc12}$ as $n\to1$. They certainly do vanish
as $n\to1$, since then the branch-point twist field becomes the
identity field. They were observed to vanish proportionally to
$n-1$ in \cite{entropy}. We first write the two-point function in
the two-particle approximation: \beq\label{pss}
    \langle \mathcal{T}(r) \tilde{\mathcal{T}}(0)\rangle \approx
    \langle\mathcal{T}\rangle^2 \lt(1 + {\mbox{1-part.}\atop \mbox{terms}} + \frc{n}{8\pi ^2} \sum_{\alpha,\beta=1}^\ell
    \int\limits_{-\infty}^{\infty } \int\limits_{-\infty }^{\infty }d\theta_{1}d\theta _{2} f_{\alpha,\beta}(\theta_{12},n)\,
    e^{-r(m_\alpha\cosh\theta_1 + m_\beta\cosh \theta_2)} \rt)
\eeq
where
\begin{eqnarray} \label{deff}
 {\langle\mathcal{T}\rangle^{2}}f_{\alpha,\beta}(\theta,n) &=&
 \sum_{j=1}^{n}\left|F_{2}^{\mathcal{T} |(\alpha,1)(\beta,j)}(\theta,n)\right|^2 \\
   &=& \left|F_{2}^{\mathcal{T} |\alpha\beta}(\theta,n) \right|^2 +
  \sum_{j=1}^{n-1} \left|F_{2}^{\mathcal{T} |\alpha\beta}(-\theta+2\pi ij,n)\right|^2.
  \nonumber
\end{eqnarray}
There is no contribution from the one-particle terms when we take
the derivative  with respect to $n$ and evaluate it at $n=1$,
since they are squares of (analytically continued) one-particle form factors, which are
$\theta$-independent and vanish faster than $n-1$ as
$n\to1$ by assumption.

As for the function $f_{\alpha,\beta}(\theta,n)$, coming from
two-particle form factors, let us denote by
$\t{f}_{\alpha,\beta}(\theta,n)$ its analytic continuation from
$n=1,2,\ldots$ to $n\in[1,\infty)$. The analytic continuation in
$n$ of the form factors themselves, for fixed rapidity, is natural
from the geometrical picture of conical singularities with angle
$2\pi n$. For any $n$ real and positive, form factors should have
kinematic poles at $\theta=i\pi$ and $\theta=2i\pi n-i\pi$
representing particles going in a straight line on either sides of
the conical singularity. They also possibly have bound state poles
as described above corresponding to bound states forming on either
side, and they satisfy all other properties stated above, now with
$n$ real and positive. For technical reasons, we must assume that
the residues at bound state poles are not diverging as $n\to1$. It
is natural to expect that for any fixed rapidity, the resulting
form factors do not ``oscillate'' as functions of $n$. More
precisely, that they have definite convexity, at least for large
enough $n$. As for the sum over $j$ in (\ref{deff}), it should be
understood, when multiplied by $n$ like in (\ref{pss}), as the
total two-particle contribution with particles allowed to cover
the region all around the conical singularities. Intuitively, it
should be smooth and should not present any oscillations either in
$n$, since the space around the conical singularities just
increases linearly with $n$. Hence it should have definite
convexity, at least for $n$ large enough (where parallel incoming
straight trajectories just to the right and to the left of the
conical singularity are far apart past it). In fact, a natural and
more constraining requirement is that no oscillatory terms are
present in the large-$n$ expansion. These properties may fix
uniquely the analytic continuation in $n$.

A smooth function of $n>1$ can be obtained easily from the sum
over $j$. Methods used in \cite{entropy} lead to the fact that in
the limit $n\to1$, the only possible contribution to the
derivative w.r.t. $n$ is from the collision of the kinematic poles
in $F_2^{{\cal T}|\alpha\beta}(\theta,n)$ when
$\alpha=\bar{\beta}$. This uses the assumption that form factors
vanish faster than $(n-1)^{\frc12}$, and the fact that these are
the only singularities that can collide as $n\to1$. Hence the
contribution can be evaluated exactly by extracting these
kinematic poles. For completeness and clarity, let us provide here
the explicit analytic continuation of the sum over $j$, following
\cite{entropy} and emphasising the general principles used. We
write \beq
    \sum_{j=1}^{n-1} \left|F_{2}^{\mathcal{T} |\alpha\beta}(-\theta+2\pi ij,n)\right|^2 =
    \frc1{2i}\oint ds \,g(s) \cot\pi s - Q \label{als}
\eeq
where
\beq
    g(s) = F_{2}^{\mathcal{T} |\alpha\beta}(-\theta+2\pi is,n) (F_{2}^{\mathcal{T} |\alpha\beta})^*(-\theta-2\pi is,n)~.
\eeq
The contour ranges from ${\rm Re}(s)=\ep$ to ${\rm Re}(s)=n-\ep$ encircling counter-clockwise the segment $s\in[\ep,n-\ep]$ for some $0<\ep<1/2$ near enough to $1/2$ so that no bound state singularity can be encircled. If the form factors at infinite rapidities vanish sufficiently fast (in all examples calculated, they vanish exponentially, which is sufficient), then the contour can be widened up to ${\rm Im}(s) = \pm \infty$, since the contribution at ${\rm Im}(s) = \pm\infty$ vanishes. The quantity $Q$ is the sum of the residues from the kinematic poles in the product of form factors themselves:
\beqa
    Q &=& \sum_{\mbox{\small kinematic poles }\t{s}}
        {\rm Res}_{s=\t{s}} \lt(F_{2}^{\mathcal{T} |\alpha\beta}(-\theta+2\pi is,n) (F_{2}^{\mathcal{T} |\alpha\beta})^*(-\theta-2\pi is,n) \rt)
        \pi \cot\pi \t{s} \n
    &=& \delta_{\b\alpha,\beta} \tan\lt(\frc\theta2\rt) {\rm Im}\lt(F_{2}^{\mathcal{T} |\alpha\b\alpha}(-2\theta+i\pi) -
        F_{2}^{\mathcal{T} |\alpha\b\alpha}(-2\theta+2i\pi n -i\pi)\rt)~. \label{exQ}
\eeqa
Using the $n$-periodicity of $\cot\pi s$, the part of the integration at ${\rm Re}(s) = n-\ep$ can be re-written, and the contour integral in (\ref{als}) becomes
\beq\label{alss}
    \frc12\int_{-\infty}^{\infty} dy \lt(g(n+i y-\ep) \cot\pi (iy-\ep) - g(iy+\ep)\cot\pi(iy+\ep)\rt)
\eeq This expression is analytic in $n$; as $n$ is varied
continuously down to $n=1$,  no pole crosses the $y$ integration
line. The expression for $Q$ is analytic in $n$ for $n>1$ as well.

Note that we used the function $\pi\cot\pi s$, which is a function
that remains unchanged under $s\mapsto s+n$ for any integer $n$
and that has simple poles with equal residues, in the band ${\rm
Re}(s)\in(0,n)$, only on the integers $1,2,\ldots,n-1$. The poles
were used to reproduce the correct sum, and periodicity was used
to cure the part of the integration contour at ${\rm Re}(s) =
n-\ep$, which otherwise would have crossed poles as $n$ is varied
continuously. However, we wish to emphasise that no periodicity
property of the function $g(s)$ (related to the form factors)
itself is used, contrary to the derivation presented in
\cite{entropy}. The function $g(s)$ does not cause problems since
as $n$ is varied, its poles move accordingly. Could we replace
$\pi\cot\pi s$ by another function $u(s)$? If we require that
$u(s)$ do not depend explicitly on $n$ in addition to the
properties above, then it is uniquely fixed to $\pi\cot \pi s$.
Relaxing periodicity to, for instance, $u(s+1) = v(s)u(s)$ would
impose $v(s)=1$ for $s$ integer (in order to have the correct
residues), and conditions on convergence at infinity would require
$v(s)=1$. Note also that the expression (\ref{alss}) does not
depend on $\ep$ even for $n$ non-integer.

The integral (\ref{alss}) have definite convexity as function of
$n$, at least for large enough $n$. This follows from expected
convexity properties of the form factors themselves. The same
holds for the residues $Q$. Integration over $\theta$ should
preserve these properties for large enough $n$, and it seems
natural to expect that the full large-$n$ expansion does not
contain oscillatory terms. Of course, a more complete analysis of
the $n$-dependence of $Q$ and of the integral (\ref{alss}) would
be useful.

It is clear that the integral in (\ref{alss}) vanishes when
$n\to1$ faster then $n-1$ by our assumption, and that $Q$ vanishes
at least like $(n-1)^{3/2}$ for any $\theta\neq0$. The only
possible obstruction in $Q$ is when $\theta\to0$, due to the
kinematic pole of the first term in (\ref{exQ}) at argument
$(2n-1)i\pi$, and the kinematic pole of the second term at
argument $i\pi$. Extracting these poles gives, for $n\to1$ and
$\theta\to0$, \beq
    \t{f}_{\alpha,\beta}(\theta,n) = \delta_{\b\alpha,\beta} \t{f}_{\alpha,\b\alpha}(0,1) \lt( \frc{i\pi (n-1)}{2(\theta+i\pi(n-1))} - \frc{i\pi (n-1)}{2(\theta-i\pi(n-1))}\rt)
\eeq
with
\beq
    \t{f}_{\alpha,\b\alpha}(0,1) = \frc12~.
\eeq Here and below we adopt the convention that
$\t{f}_{\alpha,\b\alpha}(0,1)=\lim_{n\to1}
\t{f}_{\alpha,\b\alpha}(0,n)$. This immediately gives
\begin{equation}
    \lt(\frc{\p}{\p n} \t{f}_{\alpha,\beta}(\theta,n) \rt)_{n=1} = \frc{\pi^2}2 \delta(\theta) \delta_{\b\alpha,\beta}~.
\end{equation}
This has a non-zero measure at $\theta=0$, and proves (\ref{main}).

As we mentioned above, there are two main requirements for the
argument to be valid: first, an appropriate analytic continuation
has to be taken, in agreement with the intuition from conical
singularities, and second, the analytically continued one- and
two-particle form factors have to vanish, for any real rapidity,
faster than $(n-1)^{\frc12}$ as $n\to 1$. As we said above, for
the latter requirement, in the cases constructed in
\cite{entropy}, it was observed that form factors vanish like
$n-1$, and this is what is expected in general. Below we provide
further evidence for both requirements in the case of the
two-particle form factors by constructing them explicitly in the
sine-Gordon model. In particular, we verify numerically the
striking fact that $\t{f}_{\alpha,\b\alpha}(0,1) = \frc12$ is in
agreement with $n\t{f}_{\alpha,\b\alpha}(0,n)$ having no
oscillatory terms in its large-$n$ expansion. We also find that it
diverges like $n$ as $n\to\infty$ in the super-renormalisable
case, and like $n\log n$ in the marginally renormalisable case.

\section{The sine-Gordon case}

Let us now consider the specific sine-Gordon case. This model can be defined by the lagrangian
\beq
    {\cal L} = \frc12(\p_0\varphi)^2-\frc12(\p_1\varphi)^2 + \mu\cos(\beta\varphi)~,
\eeq where $\mu$ has scaling dimension $2\beta^2-2$ and $\beta$
does not renormalise. We will use the variable \beq
    \nu = \frc{\beta^2}{8\pi-\beta^2}~.
\eeq For $\nu\ge1$, the spectrum of the model is known to be
composed of two particles with  equal masses, which we will label
by $+$ and $-$, representing the quantised version of soliton and
anti-soliton in the classical theory. For simplicity we will
consider this region only. For $\nu<\infty$, the model is
super-renormalisable. At $\nu=1$, the particles are free with
fermionic statistics, and the model is equivalent to a massive
Dirac theory. The model with $\nu>1$ can be seen as the massive
Thirring model, a perturbation of the massive Dirac theory that
preserves the $U(1)$ symmetry. From this viewpoint, the asymptotic
particles are the positively and negatively charged version of the
same particle. As $\nu\to\infty$, the model specialises to (a
sector of) the $SU(2)$-Thirring model, which is marginally
renormalisable.

The scattering of soliton with anti-soliton is non-diagonal. The
scattering matrix is given by \cite{Zamolodchikov:1977yy}
\begin{equation}\label{bc}
S_{+-}^{+-
}(\theta)=\frac{\sinh\left(\frac{\theta}{\nu}\right)}{\sinh\left(\frac{i\pi-\theta}{\nu}\right)}
S_{++}^{++}(\theta), \qquad S_{+-}^{-+
}(\theta)=\frac{\sinh\left(\frac{i
\pi}{\nu}\right)}{\sinh\left(\frac{i\pi-\theta}{\nu}\right)}
S_{++}^{++}(\theta),
\end{equation}
with
\begin{equation}\label{spp}
 S_{++}^{++}(\theta)= S_{--}^{--}(\theta)=-\exp\left[\int_{0}^{\infty}\frac{dt}{t}
\frac{\sinh\left(\frac{t(1-\nu)}{2}\right){\sinh\left(
\frac{t\theta}{i\pi}\right)} }{\sinh(t \nu/2)\cosh(t/2)}\right]~.
\end{equation}

Since branch-point twist fields have zero $U(1)$ charge, their one-particle form factors vanish. On the other hand, the relations (\ref{maineq}) give the following system of coupled equations for their two-particle form factors:
\begin{eqnarray}
F_{2}^{\mathcal{T}|+ -}(\theta, n)&=&\sum_{\gamma, \delta=
\pm}F_{2}^{\mathcal{T}|\gamma \delta }(-\theta, n)S_{-+ }^{\delta
\gamma}(\theta)= F_{2}^{\mathcal{T}|
- +}(2\pi i n-\theta, n),\\
F_{2}^{\mathcal{T}|- +}(\theta, n)&=&\sum_{\gamma, \delta=
\pm}F_{2}^{\mathcal{T}|\gamma \delta }(-\theta, n)S_{+- }^{\delta
\gamma}(\theta)= F_{2}^{\mathcal{T}|+-}(2\pi i n-\theta, n).
\end{eqnarray}
These equations can be diagonalised by introducing the linear combinations $F_{2}^{\mathcal{T}|+-}(\theta, n) \pm F_{2}^{\mathcal{T}|-+}(\theta, n)$. However, the branch-point twist fields are also invariant under charge conjugation $+\leftrightarrow -$. Indeed, this is a symmetry of the action, which keeps unchanged partition functions on Riemann surfaces, and branch-point twist fields are associated to such partition functions. Hence $F_{2}^{\mathcal{T}|+-}(\theta, n)=F_{2}^{\mathcal{T}|-+}(\theta,n)$ and we are left with
\begin{equation}\label{fpm}
    F_{2}^{\mathcal{T}|+-}(\theta, n) = \left(S_{+-}^{+- }(\theta) + S_{+- }^{-+
}(\theta)\right) F_{2}^{\mathcal{T}|+-}(-\theta, n) =F_{2}^{\mathcal{T}|+-}(2\pi i n-\theta, n)~.
\end{equation}
It is possible to find integral representations for the combination of $S$-matrix elements above,
\begin{eqnarray}\label{bpc}
S_{+-}^{+- }(\theta) + S_{+- }^{-+ }(\theta) &=&
\frac{\sin\left(\frac{\pi-i\theta}{2
\nu}\right)}{\sin\left(\frac{\pi+i\theta}{2 \nu}\right)}
S_{++}^{++}(\theta)\\&=&-
\exp\left[\int_{0}^{\infty}\frac{dt}{t}\frac{2\sinh\left(\frac{t(\nu-1)}{2}\right)\cosh\left(\frac{t(\nu-2)}{2}\right)
\sinh\left(\frac{t\theta}{i \pi }\right)}{\sinh(\nu t)
\cosh(t/2)}\right]~,
\end{eqnarray}
where we used
\begin{equation}\label{a}
    \frac{\sin\frac{\pi}{2}(a+x)}{\sin\frac{\pi}{2}(a-x)}=\exp\left[\int_{0}^{\infty}\frac{dt}{t}\frac{2 \sinh t(1-a)\sinh(tx)}{\sinh t}\right],\quad \text{for}
    \quad 0<a<1.
\end{equation}
Following \cite{entropy}, it is now straightforward to find a minimal solution to
(\ref{fpm}) (that is analytic for ${\rm Im}(\theta) \in [0,2\pi n]$) up to normalisation:
\begin{equation}
    F_2^{\text{min}}(\theta,n)=-i \sinh\left(\frac{\theta}{2n}\right)\exp
    \left[\int_{0}^{\infty}\frac{dt}{t}
    \frac{2\sinh\left(\frac{t(\nu-1)}{2}\right)\cosh\left(\frac{t(\nu-2)}{2}\right)
    \sin^2 \frac{i t}{2}\left(n-\frac{\theta}{i \pi}\right)}{\sinh(n t) \sinh (t\nu) \cosh(t/2)}
    \right]~.
\end{equation}
The two particle form factor $F_{2}^{\mathcal{T}|+-}(\theta, n)$ can be fixed by including the right pole
structure as in \cite{entropy}:
\begin{equation}
F_{2}^{\mathcal{T}|+-}(\theta, n)=\frac{ \langle
\mathcal{T}\rangle \sin\left(\frac{\pi}{n}\right)}{2 n
\sinh\left(\frac{ i \pi-\theta}{2n}\right)\sinh\left(\frac{i\pi
+\theta}{2n}\right)} \frac{ F_2^{\text{min}}(\theta,n)}{
F_2^{\text{min}}(i \pi,n)},\label{full}
\end{equation}
where the normalization has been chosen so that the kinematic
residue equation gives
\begin{equation}
    F_{0}^{\mathcal{T}}=\langle \mathcal{T}\rangle.
\end{equation}

\section{Identifying the ultraviolet conformal dimension}

It is interesting to  check the solutions above for consistency. A
possible consistency check consists of analyzing the
short-distance behaviour of two-point functions involving the
twist field and compare that behaviour to the one expected from
conformal field theory predictions. In particular, one can look at
the two-point function of the twist field with the trace of the
stress-energy tensor, $\Theta$. It was found in \cite{DSC} that
\begin{equation}\label{delta}
    \Delta^{\mathcal{T}} = \Delta^{\tilde{\mathcal{T}}}=-\frac{1}{2\langle \mathcal{T}
    \rangle}\int_{0}^{\infty} r
    \left\langle \Theta(r) \mathcal{T}(0)  \right\rangle dr
\end{equation}
(where the integration is on a space-like ray). This formula is
known as the $\Delta$-sum rule. The first equality, expected from
CFT, holds from the $\Delta$-sum rule thanks to the fact that
$\Theta$ commute with $\mathcal{T}$ and that $\Theta^\dag=\Theta$.
Following the derivations in \cite{Calabrese:2004eu,entropy} the
conformal dimension is expected to be
$\Delta^{\mathcal{T}}=d_n/2$, where $d_n$ was given in
(\ref{dim}).

As explained in detail in \cite{entropy} we can now employ the
expansion of the two-point function in terms of two particle form
factors and obtain
\begin{eqnarray}
\Delta^{\mathcal{T}}  &\approx & -\frac{n}{\left\langle
\mathcal{T}\right\rangle } \int\limits_{-\infty }^{\infty
}\int\limits_{-\infty }^{\infty }\frac{d\theta _{1} d\theta _{2}
F_{2}^{\Theta |+-}(\theta _{12}) F_{2}^{\mathcal{T}
|+-}(\theta_{12},n)^{*}}{2 (2\pi )^{2} m^2 \left( \cosh \theta
_{1} + \cosh \theta_2 \right) ^{2}}\n &=& -\frac{n}{ 8 \pi^2 m^2
\left\langle \mathcal{T}\right\rangle } \int\limits_{0}^{\infty
}\frac{d\theta F_{2}^{\Theta |+-}(\theta) F_{2}^{\mathcal{T}
|+-}(\theta,n)^{*}}{\cosh^2 (\theta/2)}, \label{dcorr2}
\end{eqnarray}
where we have used the fact that $F_{2}^{\Theta |+-}(\theta
_{12})=F_{2}^{\Theta |-+}(\theta _{12})$ and $F_{2}^{\Theta |\pm
\pm}(\theta _{12})=0$.

The two particle form factors of $\Theta$ can be obtained from
results already known in the literature. In particular, by
employing the results obtained in \cite{Luky1,Luky2} for the form
factors of exponential fields and the relation between the
sine-Gordon coupling constant and the soliton mass we find:
\begin{equation}\label{ft}
   F_{2}^{\Theta |+-}(\theta )=
   \frac{ 2\pi i m^2}{\nu}\frac{G(\theta) \cosh\frac{\theta}{2}}{\sinh\left(\frac{
   i \pi-\theta}{2\nu}\right)},
\end{equation}
where
\begin{equation}\label{g}
    G(\theta)=-i \sinh\frac{\theta}{2}\exp \left[\int_{0}^{\infty} \frac{dt}{t} \frac{\sinh t(1-\nu)
    \sin^2 i t \left(1-\frac{\theta}{i\pi}\right)}{\sinh(2t)\cosh t \sinh(t
    \nu)}\right].
\end{equation}
A table of values of the integral (\ref{dcorr2}) for different
values of $n$ and $\nu$ is presented below. The values given in
the top row in brackets are the exact values of the dimension, as
predicted by CFT. Recall that we assumed the non-existence of
bound states, which is why we only consider the case $\nu >1$.
\begin{center}
\begin{tabular}{|l|l|l|l|l|}
\hline & $n=2$ (0.0625) & $n=3$ (0.1111) & $n=4$ (0.1563) & $n=5$ (0.2)\\
\hline
 $\nu=1.1$ & 0.0627 & 0.1115 & 0.1568 &0.2008  \\ \hline
 $\nu=1.6$ & 0.0628 & 0.1118 & 0.1573 &0.2013  \\\hline
 $\nu=2.1$ & 0.0627 & 0.1115 & 0.1570 &0.2010  \\\hline
 $\nu=2.6$ & 0.0626 & 0.1113 & 0.1566 &0.2003 \\\hline
 $\nu=3.1$ & 0.0625 & 0.1111 & 0.1562 &0.1999  \\\hline
 $\nu=3.6$ & 0.0624 & 0.1111 & 0.1560 &0.1995  \\\hline
\end{tabular}
\end{center}
\begin{center}
\begin{tabular}{|l|l|l|l|l|l|}
\hline & $n=6$ (0.2431)& $n=7$ (0.2857)& $n=8$ (0.3281)& $n=9$ (0.3704) & $n=10$ (0.4125)\\
\hline
 $\nu=1.1$& 0.2440  &0.2868 &0.3293& 0.3717 &0.4139  \\ \hline
 $\nu=1.6$ &0.2447  &0.2876  &0.3303& 0.3728 &0.4152 \\\hline
 $\nu=2.1$ &0.2443  &0.2871  &0.3297&0.3722 &0.4145  \\\hline
 $\nu=2.6$ &0.2434  &0.2861  &0.3287&0.3707 &0.4129  \\\hline
 $\nu=3.1$ &0.2428  &0.2854  &0.3277&0.3698 &0.4119  \\\hline
 $\nu=3.6$ &0.2424  &0.2848  &0.3270&0.3690 &0.4109 \\\hline
\end{tabular}
\end{center}
As we can see, all values are extremely close to the exact ones,
even in the two particle approximation. This provides a
consistency check for our form factor (\ref{full}). A further
check concerns the correct normalization of the form
factor (\ref{ft}). Indeed the normalization employed here
corresponds to $ F_{2}^{\Theta |+-}(i \pi)=2 \pi m^2$ and differs
from that in other places in the literature \cite{Luky1}. A
consistency check for this normalization consists of
extracting the central charge $c=1$ of the underlying CFT by
employing Zamolodchikov's $c$-theorem \cite{Zamc}:
\begin{eqnarray}
  c = \frac{3}{2}\int_{-\infty}^{\infty} dr r^3 \left\langle \Theta(r) \Theta(0)
  \right\rangle.
\end{eqnarray}
Employing again the expansion of the two point function in terms
of form factors, changing variables and performing one integral as
in (\ref{dcorr2}) we obtain
\begin{eqnarray}
  c \approx
  \frac{3}{8 \pi^2}\int_{0}^{\infty} \frac{\left|F_{2}^{\Theta |+-}(\theta
    )\right|^2}{\cosh^4 (\theta/2)}d \theta.
\end{eqnarray}
This integral can be easily evaluated numerically for different
values of $\nu$ and gives
\begin{center}
\begin{tabular}{|l|l|l|l|l|l|}
\hline & $\nu=1.1$& $\nu=2.1$& $\nu=3.1$& $\nu=4.1$& $\nu=5.1$\\
\hline
 $c$& 0.9999& 0.9918 &0.9888 &0.9877 & 0.9874\\ \hline
\end{tabular}
\end{center}

\section{Computation of $\tilde{f}_{\pm\mp}(0,n)$: Numeric and analytic results}
\indent In this section we wish to study the behaviour of the
function $\tilde{f}_{\pm \mp}(0,n)$ (see (\ref{deff})) in detail
and, most importantly, show that  $\tilde{f}_{\pm \mp}(0,1)=1/2$
as expected. Let us introduce the notation $\t{f}(n) \equiv
\t{f}_{+-}(0,n) = \t{f}_{-+}(0,n)$. This function, by definition,
is the natural analytic continuation of $f_{+-}(0,n)$ from
$n=2,3,\ldots$ to $n\in[1,\infty)$, that is with the prescription
of smoothness and monotony as described in paragraph 2.2.
We will obtain the form of its full
large-$n$ expansion, with explicit first few coefficients, showing that it grows proportionally to $n$
for any finite $\nu$, and to $n\log n$ for $\nu=\infty$. This
large-$n$ expansion is interesting in its own right,  and turns
out to differ quite dramatically from the $1/n$ expansion found
for the sinh-Gordon model \cite{entropy}. We will then perform a
numerical study to confirm it, and most importantly, verify the
value $\tilde{f}(1)=1/2$.

Let us commence by our analytical considerations. Note that the
first term of (\ref{deff}) is zero at $\theta=0$. For the summand,
the second identity of (\ref{fpm}) implies that
$F_{2}^{\mathcal{T}|+-}(2 \pi i j, n)=F_{2}^{\mathcal{T}|+-}(2 \pi
i (n-j), n)$ for $j=1, \ldots, n-1$. In addition, for any fixed
value of $n$ the value of the function is largest for smaller
values of $j$ and decreases quickly as $j$ approaches $[n/2]$.
This behaviour is particularly extreme for $n$ very large. In that
case, only for $j \ll n$ does the function above have
non-negligible values. Therefore, the value of $\t{f}(n)$ for $n
\rightarrow \infty$ can be obtained by replacing the sum
$\sum_{j=1}^{n-1} |F_{2}^{\mathcal{T}|+-}(2 \pi i j, n)|^2$ by
\begin{equation}\label{sum}
2 \sum_{j=1}^{\infty}
\lim_{n \rightarrow \infty} |F_{2}^{\mathcal{T}|+-}(2 \pi i j, n)|^2 = \tilde{f}(\infty).
\end{equation}
This can be written as
\begin{equation}
    \tilde{f}(\infty)=\frac{2}{b(1/2)^2} \sum_{j=1}^{\infty} a(j)^2
    b(j)^2,\label{ab}
\end{equation}
where
\begin{equation}\label{aj}
a(j)=\lim_{n \rightarrow \infty} a(j,n)= \frac{2 (2j)^{\frac{3}{2}
- \frac{1}{2\nu}}}{\pi(1 - 4j^2) },
\end{equation}
with
\begin{equation}\label{ajn}
    a(j,n)=\frac{
      \sin (\frac{\pi }{n})\,
      {\sin (\frac{j\,\pi }{n})}^{\frac{3}{2} - \frac{1}{2\nu}}
      }{2n\sin(\frac{(1-2j)\pi }{2\,n})\sin (\frac{(1+2j)\pi }{2n})\sin (\frac{\pi }{2n})^{\frac{3}{2} - \frac{1}{2\nu}}
      },
\end{equation}
 and
\begin{eqnarray}\label{b}
    b(j)&=&\lim_{n \rightarrow \infty} b(j,n)=\lim_{n \rightarrow \infty} {F_2^{\text{min}}(2 \pi i j,n)}
    {\left[\sin\left(\frac{j \pi}{n}\right)\right]^{\frac{1}{2 \nu}-\frac{3}{2}}
    } \nonumber \\
    &=& \exp
    \left[-\int_{0}^{\infty}\frac{dt}{2t}\left(
    \frac{2\sinh\left(\frac{t(\nu-1)}{2}\right)\cosh\left(\frac{t(\nu-2)}{2}\right)
   }{\sinh (t\nu) \cosh(t/2)}-
    \frac{\nu-1}{\nu}\right)e^{-2tj}
    \right],
\end{eqnarray}
with
\begin{equation}\label{bjn}
    b(j,n)=\exp
    \left[\int_{0}^{\infty}\frac{dt}{t}\left(
    \frac{2\sinh\left(\frac{t(\nu-1)}{2}\right)\cosh\left(\frac{t(\nu-2)}{2}\right)
   }{\sinh (t\nu) \cosh(t/2)}-
    \frac{\nu-1}{\nu}\right)\frac{\sin^2 \frac{i t}{2}\left(n-2j\right) }{ \sinh(n
    t)}
    \right].
\end{equation}
Notice that the factor $\sin\left(\frac{j
\pi}{n}\right)^{\frac{1}{2 \nu}-\frac{3}{2}}$ in $a(j)$ is
cancelled by a similar factor in $b(j)$. They have been introduced
in order to guarantee that the integral in $b(j)$ is convergent as
$t \rightarrow 0$. Formula (\ref{ab}) holds only for $\nu$ finite;
we will come back to the case $\nu=\infty$ below. Evaluating
(\ref{ab}) numerically we obtain the values listed in the table at
the end of this section.

However, contrarily to the sinh-Gordon example studied in
\cite{entropy}, the large-$n$ corrections to this linear behaviour
are not just of $1/n$ type but depend on the particular value of
$\nu$. The form of these corrections can be determined by some
analysis. Notice that
\begin{equation}
    \tilde{f}(n)=\lt\{\ba{ll} \frac{2}{b(1/2,n)^2} \sum_{j=1}^{n/2} a(j,n)^2
    b(j,n)^2 - \frc1{b(1/2,n)^2} a(n/2,n)^2 b(n/2,n)^2 & \mbox{$n$ even} \z
    \frac{2}{b(1/2,n)^2} \sum_{j=1}^{(n-1)/2} a(j,n)^2
    b(j,n)^2 & \mbox{$n$ odd} \ea \rt.
    \label{abc}
\end{equation}
Let us first analyse $b(j,n)$ and $a(j,n)$ independently. It is trivial to see that
$\lim_{j\rightarrow \infty} b(j)^2=1$. More precisely, it has a large-$j$ asymptotic expansion of the form
\begin{equation}\label{bj}
    b(j)^2 = \sum_{k=0}^{\infty} b_k \,(2j)^{-2k} +
    O(j^{-\infty}), \qquad \text{with} \qquad b_0=1.
\end{equation}
 where we use powers of $2j$ instead of $j$ for later convenience.
This formula is obtained from a small-$t$ expansion of the
coefficient of $e^{-2tj}$ in the integrand in the last line of
(\ref{b}). The corrections for finite but large $n$ can be
obtained by looking at $b(j,n)/b(j)$. Expanding the resulting
integrand in $t$ while keeping intact the factor $\sin^2
\frc{it}2(n-2j)/\sinh(nt)$ and the factor $e^{-2tj}$, then by
integrating this expansion term by term, we find
 \beq\label{corrb}
    \frc{b(j,n)}{b(j)} = 1+\frc{(\nu-1)(7\nu-5)}{96\nu n^2}\lt(\zeta\lt(2,1-\frc{j}{n}\rt)+\zeta\lt(2,1+\frc{j}{n}\rt)-\pi^2\rt) + O\lt(\frc{1}{n^4}\rt)
\eeq where $\zeta(z,a) = \sum_{k=0}^{\infty} (k+a)^{-z}$ is the
generalized Riemann zeta function. The full series is just the
exponential of a series that has terms in $n^{-2k}$ for
$k=1,2,\ldots$, with coefficients that are functions of $j/n$,
linear in the functions $\zeta\lt(2k,1-\frc{j}{n}\rt) +
\zeta\lt(2k,1+\frc{j}{n}\rt)$\footnote{Note that all such
functions can be written in terms of trigonometric functions, for
instance $\zeta(2,1+z)+\zeta(2,1-z) = \pi^2/\sin^2\pi z - 1/z^2$,
but the analytic structure is clearer using the generalised zeta
functions.}. This series expansion is convergent for all
$j=1,\ldots,n/2$.

As for the function $a(j)$, its square has a large-$j$
series expansion
\begin{equation}\label{aaj}
a(j)^2 = \frac{4}{\pi^2 (2j)^{1+\frac{1}{\nu}}}
\sum_{k=0}^{\infty} a_k j^{-2k},\qquad \text{with}\qquad a_0=1.
\end{equation}
 The square of its corrected
form $a(j,n)$ can be expanded at large $n$ as follows:
\beq\label{corra}
    a(j,n)^2 = a(j)^2 + \frc{(2j)^{3-\frc1\nu}}{(4j^2-1)n^2} \sum_{k=0}^\infty p_{k}(j^2)n^{-2k}
\eeq
where $p_k(j^2)$ is a polynomial of order $k$ in $j^2$. Again, this series expansion is convergent for all $j=1,\ldots,n/2$. In fact, it converges for all values of $n$ such
that $|n|>|j\pm 1/2|$ and $|n|>1/2$.

We now consider the large-$n$ expansion of (\ref{abc}). We will do
explicitly the case where $n$ is even,  and comment about the
agreement with the case where $n$ is odd afterwards. First, the
subtraction $- \frc1{b(1/2,n)^2} a(n/2,n)^2 b(n/2,n)^2$ for $n$
even can be seen, from the analysis above, to be \beq\label{wps}
    -\frc1{b(1/2)^2}\lt(\frc2{\pi}\rt)^{1-\frc1\nu} n^{-1-\frc1\nu} (1+n^{-2}[[n^{-2}]])
\eeq
where here and below we use the notation
\beq
    [[x]] = \mbox{some series in non-negative integer powers of $x$.}
\eeq Let us then consider only the term which is a sum, and
subtract for convenience the value $b(1/2)^2/b(1/2,n)^2
\t{f}(\infty)$, which is just $[[n^{-2}]]$. We are left with
\begin{equation}
  \frac{ 2}{b(1/2,n)^2}\lt(\sum_{j=1}^{n/2} a(j,n)^2 b(j,n)^2 -
\sum_{j=1}^\infty a(j)^2 b(j)^2\rt).
\end{equation}
The pre-factor $2/b(1/2,n)^2$ is also $[[n^{-2}]]$, so we will
consider the sum without pre-factor. We will consider three parts:
\begin{eqnarray}
\text{I.} && \sum_{j=1}^{n/2} a(j)^2 b(j)^2 - \sum_{j=1}^\infty
a(j)^2
b(j)^2 = -\sum_{j=n/2+1}^{\infty} a(j)^2b(j)^2, \\
\text{II.} && \sum_{j=1}^{n/2} \{\mbox{corrections to
$a(j)^2$}\}\cdot
b(j)^2,\\
\text{III.} && \sum_{j=1}^{n/2} a(j,n)^2 \cdot\{\mbox{corrections
to $b(j)^2$}\},
\end{eqnarray}
where the correction terms are those from (\ref{corra}) and
(\ref{corrb}). In the part I, we can use the large-$j$ expansions
of $a(j)^2$ and $b(j)^2$. Consider the identity \beq\label{qfy}
    \sum_{j=n/2+1}^\infty (2j)^{-z} = 2^{-z}\zeta\lt(z,1+\frc{n}2\rt) = -n^{1-z}\lt(\frc{1}{2(1-z)} + \frc1{2n} + n^{-2} [[ n^{-2}]]
    \rt)
\eeq (here, the coefficients of the series $[[n^{-2}]]$ are
polynomials in $z$), where $\zeta(z) = \zeta(z,1)$ is the Riemann
zeta function. This implies that \beq
    {\rm I}: -\sum_{j=n/2+1}^{\infty} a(j)^2b(j)^2 = \frc{2 n^{-\frc1\nu}}{\pi^2}\lt(-\nu + \frc1n + n^{-2} [[n^{-1}]]\rt)~.
\eeq

The contributions II can be analysed using the formula
\beqa\label{uprr}
    \sum_{j=1}^{n/2} \frc{(2j)^{z}}{4j^2-1}& =& \sum_{k=0}^\infty 2^{z-2-2k}
    H_{n/2}^{(2+2k-z)}\nonumber\\
   &=& \sum_{k=0}^\infty 2^{z-2-2k} \zeta(2+2k-z) + n^{z-1}\lt( \frc{1}{2(z-1)} + \frc{1}{2n} + n^{-2}[[n^{-1}]]\rt)~.
\eeqa Here, $H_{a}^{(b)}$ is the Harmonic number of order $b$ \beq
    H_a^{(b)}=\sum_{j=1}^{a} \frac{1}{j^b} = \zeta(b) - \zeta(b,a+1)~.
\eeq Note that the infinite sum on the right-hand side of
(\ref{uprr}) is convergent for any non-integer $z$.  Also, the
coefficients in $[[n^{-1}]]$ are rational functions of $z$. Let us
write $b(j)^2 = \sum_{k'=0}^{k} b_{k'} (2j)^{-2k'} + r(j)$ where
$r(j)$ is the rest. The function $r(j)$ behaves proportionally to
$j^{-2k-2}$ as $j\to\infty$, so that the sum from 1 to $n/2$ of
the product of $r(j)$ times the $k^{\rm th}$ correction to
$a(j)^2$ in (\ref{corra}) can be extended to a sum from 1 to
$\infty$ without problems. Hence for the $k^{\rm th}$ contribution
to II we can write: \beqa\label{splf}
   && n^{-2k-2} \sum_{k'=0}^{k} \sum_{j=1}^{n/2}  b_{k'} \frc{(2j)^{3-\frc1\nu-2k'}}{4j^2-1} p_{k}(j^2)  +n^{-2k-2}
    \sum_{j=1}^\infty r(j) \frc{(2j)^{3-\frc1\nu}}{(4j^2-1)} p_{k}(j^2) \n
    && - n^{-2k-2}
    \sum_{k'=k+1}^\infty \sum_{j=n/2+1}^\infty b_{k'} \frc{(2j)^{3-\frc1\nu-2k'}}{(4j^2-1)} p_{k}(j^2) ~.
\eeqa The first term can be evaluated using (\ref{uprr}). The
second term is a contribution of the order $n^{-2k-2}$ only, and
the last term can be evaluated using the negative of (\ref{uprr})
without the infinite sum of zeta functions and is of order
$n^{-2k-2-1/\nu}$. Hence, the result is a term proportional to
$n^{-2k-2}$ plus an infinite series of the form
$n^{-\frc1\nu}[[n^{-1}]]$. For instance, the leading contributions
to this series are \beq\label{asd}
    2^{-2k} p_{k,k}n^{-\frc1\nu}\lt( \frc{1}{2(2+2k-1/\nu)} + \frc{1}{2n} +O(n^{-2})\rt)
\eeq where $p_{k,k}$ is the coefficient of $j^{2k}$ in $p_k(j^2)$.
When we consider all corrections, for all $k=0,1,2,\ldots$, we
obtain, for the coefficient of any given order $n^{-1/\nu+k'}$
with fixed $k'=0,1,2,\ldots$, an infinite sum over $k$. for
instance, these are the sums over all $k$ of (\ref{asd}) in the
cases $k'=0$ and $k'=1$. We must make sure that all these infinite
sums give finite results. By putting $n=aj$ for $a>1$ in
(\ref{corra}), a value where the series in (\ref{corra}) converges
for all $j\ge1$, we see that $\sum_{k=0}^\infty p_k(j^2)
(aj)^{-2k}<\infty$. This implies that $\sum_{k=0}^\infty p_{k,k-l}
a^{-2k}<\infty$, where $p_{k,k-l}$ is the coefficient of
$j^{2k-2l}$ in $p_k(j^2)$. Since the coefficients in $[[n^{-1}]]$
in (\ref{uprr}) are rational functions of $z$, a little analysis
of (\ref{splf}) shows that this is sufficient to prove that the
infinite sums that are the coefficients of $n^{-1/\nu+k'}$ for any
$k'=0,1,2,\ldots$ give finite results. For instance, the
contributions to $n^{-1/\nu}$ and $n^{-1/\nu-1}$ can be found from
\beqa
    h(s) \equiv \sum_{k=0}^\infty  s^{2k} p_{k,k} &=& \lim_{n\to\infty}
    \lt((a(j,n)^2-a(j)^2)\frc{(4j^2-1)n^2}{(2j)^{3-\frc1\nu}}\rt)_{j=ns}\n
    &=& \lt(\lt(\frc{\pi s}{\sin \pi s}\rt)^{1+\frc1\nu}-1\rt)\frc1{\pi^2s^2}
\eeqa by \beq
    \sum_{k=0}^\infty 2^{-2k} p_{k,k} = h(1/2) ~,\quad \sum_{k=0}^\infty \frc{2^{-2k} p_{k,k}}{2+2k-1/\nu} =
     2^{2-\frc1\nu}\int_0^{\frc12} ds\,s^{1-\frc1\nu} h(s)~.
\eeq The result is that the leading contributions for the case II
are \beq
    {\rm II}:
    n^{-2}[[n^{-2}]] + \frc{2}{\pi^2 n^{\frc1\nu}}
    \lt(\nu - \nu\lt(\frc{\pi}2\rt)^{\frc1\nu} \sqrt{\pi} \frc{\Gamma\lt(1-\frc1{2\nu}\rt)}{\Gamma\lt(\frc12-\frc1{2\nu}\rt)}
    + \frc1n \lt(\lt(\frc\pi 2\rt)^{1+\frc1\nu}-1\rt) + n^{-2}[[n^{-1}]]\rt)~.
\eeq Finally, a similar analysis can be done with the part III,
involving the correction terms of $b(j)^2$ in (\ref{corrb}), by
expanding the generalised zeta functions in powers of $j/n$.
Considerations similar to those above, with the fact that the
expansion of the generalised zeta functions is valid for
$|n|>|j|$, lead to the same structure for the large-$n$ expansion
as that of parts I and II, with an extra factor $n^{-2k}$ for each
correction term with $k=1,2,3,\ldots$. This gives: \beq
    {\rm III}: n^{-4}[[n^{-2}]] + n^{-\frc1\nu-2}[[n^{-1}]]~.
\eeq

Putting everything together, along with the subtraction (\ref{wps}) specific to the case $n$ even, we find
\beq
    n\t{f}(n) = c_0 n + \frc{c_2}{n} + \frc{c_4}{n^3} + \ldots + n^{-\frc1\nu} \lt(d_0 n + d_1 + \frc{d_2}{n} + \ldots \rt) \label{nfn}
\eeq
 where \beqa
    c_0 &=& \t{f}(\infty)\n
    d_0 &=& -\frc{\nu}{b(1/2)^2} \lt(\frc{2}\pi\rt)^{2-\frc1\nu} \sqrt{\pi} \frc{\Gamma\lt(1-\frc1{2\nu}\rt)}{\Gamma\lt(\frc12-\frc1{2\nu}\rt)}
    \label{d0}\\
    d_1 &=& 0~\nonumber.
\eeqa It is striking that although the intermediate steps of the analysis give terms
$n^{k-\frc1\nu}$ for $k=1,0,-1,-2,\ldots$, the term with $k=0$
identically vanishes for all $\nu$. In fact, one can see that the
same analysis for the case $n$ odd, starting from the second form
of (\ref{abc}), directly gives vanishing coefficients for
$k=0,-2,-4,\ldots$. For instance, changing $n\mapsto n-1$ into
(\ref{qfy}) erases the term $1/(2n)$ on the right-hand side. The
other terms computed above are also unchanged in the odd case.
Hence we conjecture that the large-$n$ expansion valid from both
$n$ even and $n$ odd is \beq\label{expnf}
    n\t{f}(n) = c_0 n + \frc{c_2}{n} + \frc{c_4}{n^3} + \ldots + n^{-\frc1\nu} \lt(d_0 n + \frc{d_2}{n} + \frc{d_4}{n^3} + \ldots \rt)
\eeq

\begin{figure}[h!]
\begin{center}
\includegraphics[width=12cm,height=8cm]{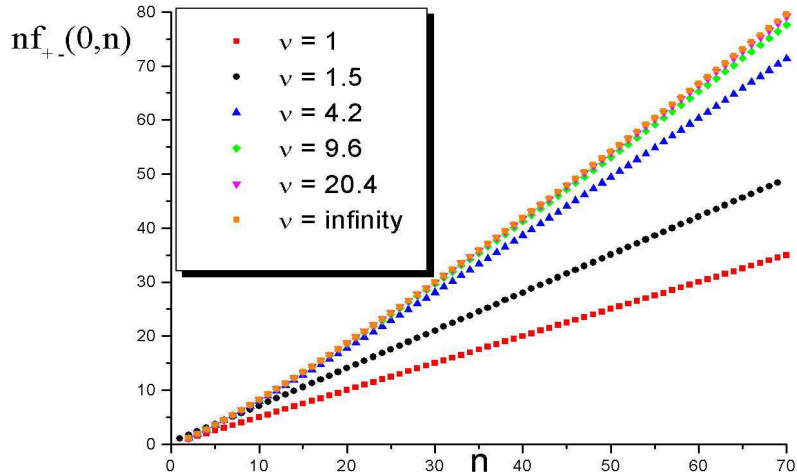}
\end{center}\vspace{-1cm}
\caption{The function $n f_{+-}(0,n)$ computed numerically for
several values of $\nu$ and $2\leq n\leq70$.} \label{f0n}
\end{figure}

Indeed a very precise numerical fit of $nf_{+-}(0,n)$ for several
values of $\nu$, is given by the function above. Let us commence
by evaluating $n\tilde{{f}}(n)$ for integers $n>1$ and several
values of $\nu$ (see Fig.~\ref{f0n}).
Notice that $\nu=1$ corresponds to the free Fermion point and,
indeed for this value of $\nu$ we recover the result obtained
analytically in \cite{entropy} for the Ising model, namely
$\tilde{f}(n)=1/2$ for all values of $n$ (in particular, also for
$n=1$). For other values of $\nu$ the large-$n$ linear behaviour
seems also apparent from Fig.~\ref{f0n} and is clearer the smaller
the values of $\nu$. In fact, for $\nu=9.6$ and $\nu=20.4$, the
correct linear behaviour can only be seen at much larger values of
$n$, because of the importance of the term in $n^{1-1/\nu}$.

Our numerical analysis has revealed that the numerical values of
$n f_{+-}(0,n)$ (especially for small $n$) are best fitted by a
function of the form:
\begin{equation}\label{F}
    F(\nu,n)=(1+n)c_0+ \tilde{c}_1+ \frac{\tilde{c}_2}{1+n}+ d_0
    (1+n)^{1-\frac{1}{\nu}}+\frac{\tilde{d}_1}{(1+n)^{\frac{1}{\nu}}}+
    \frac{\tilde{d}_2}{(1+n)^{1+\frac{1}{\nu}}},
\end{equation}
This function has the same large-$n$ behaviour as (\ref{nfn}) but
since the expansion (\ref{nfn}) is only asymptotic its working is
not guaranteed for small values of $n$ (in particular at $n=1$).
Shifting $n \rightarrow n+1$ constitutes a re-summation that allows us to consider small values of $n$.
The tables below show the values of all constants involved in
(\ref{F}) as well as the values of $c_0$ and $d_0$ as obtained by
numerically evaluating the sum (\ref{ab}) and the function
(\ref{d0}), respectively. The latter are in remarkably good
agreement with the coefficients obtained from the fit. In
addition, we give the value $F(\nu,1)$ which, for all $\nu$
considered, is compatible with the expected value
$\tilde{f}(1)=1/2$.
\begin{center}
\begin{tabular}{|c|c|c|c|c|c|}
  \hline
  $\nu$ & $\tilde{f}(\infty)$ & $c_0$ (fit) & $d_0$ (exact) & $d_0$ (fit) & $F(\nu,1)$ \\
  \hline
  1.10 & 0.548110 & 0.548100 & -0.089218 & -0.078176  & 0.496148 \\\hline
  1.50 & 0.711079 & 0.711023 & -0.326980 & -0.319507  & 0.488700 \\\hline
  4.20 & 1.400983 & 1.400000 & -1.049451 & -1.043071  & 0.483927 \\\hline
  9.60 & 2.418856 & 2.413256 & -2.042131 & -2.028276  & 0.485305 \\\hline
  20.4 & 4.266412 &  4.264102 & -3.866293 & -3.858392  & 0.489379 \\
  \hline
\end{tabular}\end{center}
\begin{center}
\begin{tabular}{|c|c|c|c|c|}
  \hline
  $\nu$ & $\tilde{c}_1$ & $\tilde{c}_2$ &$\tilde{d}_1$ & $\tilde{d}_2$  \\
  \hline
  1.10 & -0.564827 &-0.224208 & 0.267706 & 0.066029  \\\hline
  1.50 & -0.745179 &-0.189914 & 0.375405 & 0.231296  \\\hline
  4.20 & -1.613371 &-0.470833 & 1.251032 & 0.567956  \\\hline
  9.60 & -3.397554 &-3.514617 & 3.215461 & 3.431349  \\\hline
  20.4 & -5.850448 &-9.119535 & 5.615018 & 9.110232  \\
  \hline
\end{tabular}\end{center}
Concerning the values on the last table we see that, for example,
the sum $c_0+\tilde{c}_1$ which we expect to be vanishing from the
large-$n$ expansion, indeed gives relatively small numbers
compared to $c_0$ and $\tilde{c}_1$, especially for small $\nu$.
This is consistent with our working precision which is
considerably reduced for sub-leading terms. A plot of the function
$F(n,1.5)$ as well as of the numerical values of $n f_{+-}(0,n)$
for $n=2,3,...,70$ and $\nu=1.5$ from (\ref{abc}) can be seen in
Fig.~2. The difference $|n f_{+-}(0,n)-F(n,1.5)|<10^{-4}$ for all
$n=2,3,\ldots,70$. In fact, this holds for all finite values of
$\nu$ considered here.
\begin{figure}[h!]
\begin{center}
\includegraphics[width=12cm,height=8cm]{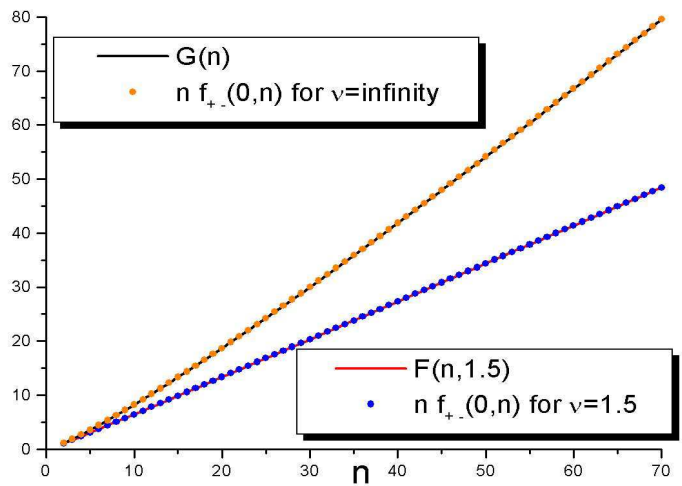}
\end{center}\vspace{-1cm}
\caption{The functions $F(n,1.5)$ and $G(n)$ and the corresponding
values of $n f_{+-}(0,n)$ for $2\leq n\leq 70$.}
\end{figure}

As we can see from the expansion, the term $n^{1-1/\nu}$ will give
a very important contribution for $\nu$ large, so that $n
\tilde{f}(n)$ should become linear in $n$ very quickly for $\nu$
close to its minimum value 1 and very slowly for large values of
$\nu$. In fact, in the limit $\nu \rightarrow \infty$, a similar
analysis as carried out above can be performed (although some
complications appear). Both $a(j,n)$ and $b(j,n)$ have
well-defined $\nu=\infty$ limit which commutes with the $n=\infty$
limit, giving \beq
    a(j)^2_{\nu=\infty} = \frc{4}{\pi^2} \frc{8j^3}{(1-4j^2)^2}~,\quad b(j)^2_{\nu=\infty} = \frc{\Gamma\lt(\frc32+j\rt)^2}{j^3\Gamma(j)^2}~.
\eeq
The leading term at large $n$, then, can be obtained from
\begin{equation}
 \sum_{j=1}^{n/2} a(j)^2_{\nu=\infty} b(j)^2_{\nu=\infty} \sim \sum_{j=1}^{n/2} \frac{ 2 }{\pi^2 j } =  \frc{2}{\pi^2} \lt(\Psi\lt(1+\frc{n}2\rt) + \gamma\rt)
\end{equation}
where $\gamma$ is Euler's number and $\Psi(z)$ is the logarithmic derivative of Euler's gamma
function $\Psi(z)=\Gamma(z)'/\Gamma(z)$. For $n$ large the
function $\Psi\left(1+n/2\right)$ behaves as
\begin{equation}
    \Psi\left(1+\frac{n}{2}\right) = \log\left(\frac{n}{2}\right) + O\lt(\frac{1}{n}\rt)~,
\end{equation}
so that the large-$n$ expansion at $\nu=\infty$ starts with
\begin{equation}
    \lim_{\nu\rightarrow \infty} n \t{f}(n) \sim \frc1{2\pi} n \log n
\end{equation}
(where we used $b(1/2)^2_{\nu=\infty} = 8/\pi$). It is quite interesting to note that this logarithmic behaviour can also be obtained from the expansion (\ref{expnf}) by
simply taking the limit $\nu\to\infty$. Both $c_0$ and $d_0$ are divergent proportionally to $\nu$, and these divergencies cancel out. The expansion in $1/\nu$
of $n^{1-1/\nu}$ then gives the correct logarithmic term. A more precise analysis of $c_0$ follows from
\beq
    \sum_{j=1}^\infty a(j)^2 b(j)^2 \stackrel{\nu\to\infty}=
        \sum_{j=1}^\infty \lt(a(j)^2 b(j)^2 - \frac{ 2 }{\pi^2 j }\rt) + \frc{2^{1-\frc1\nu}}{\pi^2} \zeta\lt(1+\frc1\nu\rt) + o(1)
\eeq
where the zeta function contains the linear divergence at large $\nu$. The first term can be evaluated exactly:
\beqa
    \sum_{j=1}^\infty \lt(a(j)^2 b(j)^2 - \frac{ 2 }{\pi^2 j }\rt) &=& \frc{2}{\pi^2} \sum_{j=1}^\infty \lt( \frc{\Gamma\lt(j-\frc12\rt)^2}{\Gamma(j)^2} - \frc1j \rt) \n
        &=& \frc{2}{\pi^2} \lim_{x\to1} (2x K(x) + \log(1-x)) \n
        &=& -\frc{4}{\pi^2} \lt(\Psi\lt(\frc12\rt) + \gamma\rt)
\eeqa
(where $K(x)$ is the complete elliptic integral of the first kind). From this, the $\nu=\infty$ limit of (\ref{expnf}) gives the conjecture
\beq
    \lim_{\nu\to\infty} 2\pi n\t{f}(n) = \lt(n + \frc{e_2}{n} + \frc{e_4}{n^3} + \ldots\rt) \log n +
    (\gamma+\log32-\log\pi )n + \frc{f_2}{n} + \frc{f_4}{n^3} + \ldots \label{ginf}
\eeq  We will now fit the values of $n f_{+-}(0,n)$ obtained
numerically to a function of the type:
\begin{equation}
    G(n)=  e_0 (1+n) \log (n+1) + \tilde{e}_1 \log(n+1)+ \frac{\tilde{e}_2 \log(n+1)}{1+n}+
    f_0 (n+1) + \tilde{f}_1 +\frac{\tilde{f}_2}{n+1}.
\end{equation}
Similarly as $F(\nu,n)$, the function $G(n)$ has a large $n$
expansion of the type (\ref{ginf}). The coefficients $e_0$ and
$f_0$ from the fit, as well as their exact values from
(\ref{ginf}) are given in the following table
\begin{center}
\begin{tabular}{|c|c|c|c|c|c|c|c|}
  \hline
  $e_0$ (exact)& $e_0$ (fit) & $f_0$ (exact) & $f_0$ (fit)  & $\tilde{e}_1$ & $\tilde{e}_2$ & $\tilde{f}_1$& $\tilde{f}_2$\\
  \hline
  0.159155 & 0.158971 & 0.461266 & 0.462474 & -0.183370
  & -0.124718 & -0.615249 & 0.248293 \\
  \hline
\end{tabular}
\end{center}
so that $G(1)=0.4838988$. Fig.~2 shows the function $G(n)$ as well
as the numerical values of $n f_{+-}(0,n)$ from (\ref{abc}). The
agreement between the two sets of values is very good. More
precisely  $|n f_{+-}(0,n)-G(n)|< 10^{-4}$ for all
$n=2,\ldots,70$. \subsection{Discussion of the large-$n$
results}\indent \\ \indent We notice from the graphs and from the
fitting functions that the value of 1/2 at $n=1$ is perfectly in
agreement with smooth and {\em convex} analytic continuations
agreeing with the given values at $n=2,3,\ldots$, in all cases
analysed. Also, the large-$n$ expansion makes it clear that no
oscillatory terms appear. We also remark that the leading
large-$n$ behaviour is linear in all super-renormalisable cases
$\nu<\infty$, and is proportional to $n\log n$ in the marginally
renormalisable case $\nu=\infty$ (this is still a non-conformal
case, where a scale appears by dimensional transmutation). This
seems to suggest that the large-$n$ behaviour is related to the
type of perturbing field, and we recall that logarithmic functions
(of the cutoff) usually appear with naively marginal
perturbations. Moreover, the full large-$n$ expansion in the
finite-$\nu$ case seems to encode some characteristics of the
perturbing field, as it also has a part with non-integer powers of
$n$ (although these powers are not in linear relation with the
dimension of the field). Also striking and pointing towards a
perturbation-theory origin of the large-$n$ behaviour is the fact
that the limit $\nu\to\infty$ can be taken directly in this
large-$n$ expansion, although it is expected to be only an
asymptotic expansion, and the limit $\nu\to\infty$ {\em cannot} be
taken in the various intermediate steps separately. We recall that
a good regularisation scheme of the sine-Gordon model at
$\nu=\infty$ is a ``non-perturbative'' dimensional regularisation,
whereby $\nu$ is taken finite (thus changing the dimension of the
perturbing field) and sent to infinity at the end of the
calculation (see for instance \cite{DL}). This leads to
logarithmic functions in a similar way to that by which $\log n$
appeared here.

\section{Conclusions and outlook}
This paper provides an extension and generalisation of the work
\cite{entropy} to integrable QFTs with backscattering. The main
conclusion following from \cite{entropy} and the present work is
that the form of the leading correction to the bi-partite entropy
at large distances of any integrable QFT is ``universal" in the
sense that it does not depend on the precise form of the
scattering matrix (in particular, whether or not it is diagonal)
but only on the mass spectrum of the model.

The argument leading to this result, developed in \cite{entropy},
uses the well-known replica trick, which involves the partition
function on Riemann surfaces with two branch points and $n$
sheets, related to the correlation function of twist fields. We
obtained the first large-distance (distance between the branch
points) correction to this partition function in the sine-Gordon
model. This correction presents an interesting large-$n$
asymptotic, which seems to be in relation with the type of
perturbing field giving the massive theory. It would be very
interesting to find a satisfying explanation for this large-$n$
behaviour, in particular in the super-renormalisable case, where
the leading behaviour is just linear.

From the replica trick, the main subtlety in obtaining the entanglement entropy is the analytic continuation in $n$,
the number of Riemann sheets. In the present
paper we attempted to provide support for the results using the geometrical picture of
conical singularities with angle $2\pi n$ instead of branch points, and appealing to some convexity properties as function of $n$.
However, a more precise statement about the uniqueness of the analytic continuation we obtained is still missing.

It would be interesting to investigate higher order corrections to
the entropy, that is, higher particle form factor contributions to
the correlation function (\ref{pss}) and, in particular, determine
whether or not they are of a similar ``universal" nature as the
leading correction already obtained. This would involve the
computation of higher particle form factors of the twist fields.
Two-particle form factors were obtained here by solving a set of form factor consistency equations,
but they could also be obtained using Lukyanov's method of angular quantisation \cite{Luk0}, as was
done in the sinh-Gordon model in \cite{entropy}. This may be quite helpful for higher particle numbers.
Also, investigation in other models may help decipher the properties of form factors of twist fields, for low-particle numbers as well.
The geometrical meaning of the field may provide simplifications. For instance, can there be non-zero one-particle form factors?

Another outstanding point is the generalisation of the approach
employed here to the computation of the entanglement entropy of
quantum systems consisting of multiply disconnected regions. Such
an extension could provide an alternative way to proving certain
natural general properties of the entanglement entropy shown to be valid in conformal field
theory \cite{Calabrese:2004eu} and recently discussed in a more general framework in \cite{veronika}.

\paragraph{Acknowledgments:}

We are grateful to John Cardy for discussions.

\small

\end{document}